\documentclass[twocolumn,amsmath,amssymb,prd]{revtex4}

\def\al{\alpha}
\def\be{\beta}
\def\ga{\gamma}
\def\de{\delta}
\def\ep{\epsilon}

\def\et{\eta}
\def\th{\theta}

\def\la{\lambda}

\def\rh{\rho}

\def\ph{\phi}

\def\ch{\chi}
\def\ps{\psi}

\def\Ga{\Gamma}

\def\fr#1#2{{{#1} \over {#2}}}
\def\half{{\textstyle{1\over 2}}}

\def\frac#1#2{{\textstyle{{#1}\over {#2}}}}

\def\lsim{\mathrel{\rlap{\lower4pt\hbox{\hskip1pt$\sim$}}
    \raise1pt\hbox{$<$}}}
\def\gsim{\mathrel{\rlap{\lower4pt\hbox{\hskip1pt$\sim$}}
    \raise1pt\hbox{$>$}}}
\def\sqr#1#2{{\vcenter{\vbox{\hrule height.#2pt
         \hbox{\vrule width.#2pt height#1pt \kern#1pt
         \vrule width.#2pt}
         \hrule height.#2pt}}}}

\def\pt#1{\phantom{#1}}

\def\vb#1#2{e_{#1}^{{\pt{#1}}#2}}
\def\ivb#1#2{e^{#1}_{{\pt{#1}}#2}}

\def\lvb#1#2{e_{#1#2}}

\def\barvb#1#2{\bar e_{#1}^{{\pt{#1}}#2}}

\def\barlvb#1#2{\bar e_{#1#2}}

\newcommand{\beq}{\begin{equation}}
\newcommand{\eeq}{\end{equation}}
\newcommand{\bea}{\begin{eqnarray}}
\newcommand{\eea}{\end{eqnarray}}
\newcommand{\bit}{\begin{itemize}}
\newcommand{\eit}{\end{itemize}}
\newcommand{\rf}[1]{(\ref{#1})}

\begin{document}

%\title{Noether identities in gravity theories with nondynamical backgrounds}
\title{Noether identities in gravity theories with nondynamical backgrounds \\
and explicit spacetime symmetry breaking}

\author{Robert Bluhm and Amar Sehic}

\affiliation{
Physics Department, Colby College,
Waterville, ME 04901, USA
}

\begin{abstract}
Gravitational effective field theories with nondynamical backgrounds 
explicitly break diffeomorphism and local Lorentz invariance.
At the same time, to maintain observer independence the action
describing these theories is required to be mathematically invariant
under general coordinate transformations and changes of local Lorentz bases.  
These opposing effects of having broken spacetime symmetries 
but invariance under mathematical observer transformations can result 
in theoretical inconsistency unless certain conditions hold.
The consistency constraints that must hold originate from 
Noether identities associated with the mathematical observer invariances in the action.
These identities are examined in detail and are used to investigate gravity 
theories with nondynamical backgrounds,
including when a St\"uckelberg approach is used.
Specific examples include gravity theories with fixed scalar or tensor backgrounds,
Einstein-Maxwell theory with a fixed external current, and massive gravity.
 \end{abstract}

%\pacs{11.30.Cp, 11.10.Ef, 04.40.Nr}

\maketitle

\section{Introduction}

The idea that Lorentz symmetry might not hold at all energy scales has received
much attention in recent years due to ideas originating from quantum gravity,
string theory, and physics at the Planck scale
\cite{ks,overviews}.
In addition, open questions related to dark energy and dark matter have led to
modified theories of gravity being proposed,
where in many cases Lorentz invariance no longer holds as an exact symmetry.
Numerous experiments searching for violations of Lorentz symmetry in
particle and gravitational interactions have been carried out.
In these tests, 
the Standard-Model Extension (SME) is widely used as the 
phenomenological framework
\cite{sme,akgrav04,rbsme},
allowing sensitivities to Lorentz violation to be expressed 
as bounds on the SME coefficients
\cite{aknr-tables}.

If local Lorentz invariance (LLI) is violated in gravitational
or particle interactions,
it is expected that the effects of this at low energy must be very small,
involving for example suppression by the Planck scale.
It is also expected on physical grounds that observer independence should 
hold even if violations of LLI occur
\cite{sme,akgrav04}.
For these reasons,
Lagrangian-based effective field theory provides a suitable approach in theoretical investigations,
where Lorentz-violating terms are incorporated in an observer-independent
action as couplings between fixed background fields and
conventional gravitational and matter fields.

In gravitational theories where violation of LLI is due to spontaneous 
symmetry breaking at a more fundamental level,
the background fields represent dynamical vacuum solutions,
which spontaneously break both 
LLI and diffeomorphism invariance.
In this case, when Nambu-Goldstone and massive excitations 
about the vacuum solution are included, 
the invariance of the action under both spacetime symmetries still holds
\cite{rbak}.
On the other hand, when a {\it nondynamical} background 
appears directly in the Lagrangian, 
the symmetry breaking is explicit.
In this case, 
there are no dynamical excitations of the background field,
and the action is not invariant under either local 
Lorentz transformations or diffeomorphisms.

When spacetime symmetry breaking is explicit
due to the presence of a nondynamical background,
it is known that potential conflicts can occur between geometrical identities,
conservation laws, and the dynamical equations of motion
\cite{akgrav04,rb15a}.
These conflicts can lead to theoretical inconsistency unless certain conditions hold.
In contrast, when the symmetry breaking is spontaneous,
the potential conflicts are evaded since the background fields are dynamical.
For these reasons most investigations using the SME assume spontaneous 
breaking of diffeomorphisms and LLI in the gravity sector.
Nonetheless, there are examples of widely studied gravitational theories 
with nondynamical backgrounds that explicitly break local diffeomorphism invariance and LLI.
Examples include Chern-Simons gravity \cite{rjsp}, 
theories with spacetime-varying couplings \cite{tvar,akrlmp03,blpr04,rb15b}, 
and massive gravity \cite{MGreviews}.

In gravitational theories with explicit breaking and nondynamical backgrounds, 
the consistency conditions that are usually verified are the equations,
$D_\mu T^{\mu\nu} = 0$ and $T^{\mu\nu} = T^{\nu\mu}$,
where $T^{\mu\nu}$ is the energy-momentum tensor obtained using either a metric or vierbein formalism.
The nondynamical background fields enter these equations as part of $T^{\mu\nu}$.

In General Relativity (GR), these consistency conditions are obtained by combining the Noether identities associated
with diffeomorphism invariance and LLI with the dynamical equations of motion for the gravitational and matter fields.
The usual interpretation of the Noether identities in GR is that they provide relations that hold off shell between 
the Euler-Lagrange expressions for the matter and gravitational fields
\cite{en18,A62,Traut62,bb2002,Obuk}.
In the case of diffeomorphism invariance, 
this leads to the result that the four equations $D_\mu T^{\mu\nu} = 0$ 
automatically hold when the matter fields are on shell.
Similarly, in GR in a vierbein treatment the Noether identities associated with LLI
provide off-shell relations involving anti-symmetric combinations of the matter and
gravitational fields and their Euler-Lagrange expressions.
On shell, these result in the six equations $T^{\mu\nu} = T^{\nu\mu}$.

In contrast, in a theory with a nondynamical background,
both diffeomorphism invariance and LLI are explicitly broken,
and the usual Noether identities stemming from these symmetries no longer apply.
Nonetheless, 
the same consistency conditions,
$D_\mu T^{\mu\nu} = 0$ and $T^{\mu\nu} = T^{\nu\mu}$,
must hold on shell even when a nondynamical background is present.
In this case, it is because these equations follow directly from Einstein's equations, 
$G^{\mu\nu} = 8 \pi G T^{\mu\nu}$, 
where $G^{\mu\nu}$ is the Einstein tensor, 
combined with the contracted Bianchi identity, $D_\mu G^{\mu\nu} = 0$, 
and the symmetry relation $G^{\mu\nu} = G^{\nu\mu}$.

At the same time,
it is important to realize that a form of Noether identities can still be found
when there is explicit spacetime symmetry breaking.
In this case, the identities stem from the requirement of observer independence, 
which provides the freedom to pick any coordinate system and any local Lorentz basis.
Because of observer independence, the action is required to be mathematically invariant under 
both general coordinate transformations and changes of local Lorentz bases.  
This leads to Noether identities that can be derived using these observer invariances of the action.
In this case, the identities involve not only the Euler-Lagrange expressions for
the dynamical fields, but also for the nondynamical background fields.
However, the nondynamical backgrounds do not have to obey their Euler-Lagrange equations.
Therefore, the usual interpretation of the Noether identities does not hold,
and a different interpretation is required when nondynamical background fields are present.

The goal of this paper is to examine the Noether identities that arise
from the requirement of observer independence in gravitational theories that 
contain a nondynamical background.
In addition, it is shown that making a detailed examination of these Noether identities 
can provide insight into the potential theoretical inconsistencies 
that can occur in a theory with explicit spacetime symmetry breaking.
For example, when the equations $D_\mu T^{\mu\nu} = 0$ or $T^{\mu\nu} = T^{\nu\mu}$
are found not to hold in a gravitational theory,
which either makes the theory inconsistent or imposes restrictions on the physical fields and/or geometry,
the underlying reasons are often not readily apparent.  
In this case, it is advantageous to look directly at the structure of the Noether identities 
associated with observer independence in order to understand how the 
inconsistency arises or is possibly evaded.

This paper is organized as follows.
The next section begins with a brief review of spacetime symmetry transformations
and the usual interpretation of the Noether identities in GR.
Since diffeomorphism invariance is manifest in a metric description while LLI remains hidden,
a vierbein formalism is used to examine LLI.
However, torsion is assumed to vanish in this investigation 
and only Riemann spacetimes are considered.
Section III discusses diffeomorphism and LLI breaking in
theories with nondynamical background fields.
The Noether identities that hold as a result of observer independence are presented,
and their interpretation is discussed in comparison to GR and theories
with spontaneous spacetime symmetry breaking.
Section IV looks at applications of these Noether identities,
including specific examples of gravity theories with nondynamical backgrounds.
Lastly, Section V gives a summary and conclusions.

\section{Spacetime Symmetries in GR}

In a Lagrangian description of GR,
diffeomorphism invariance and LLI act effectively as local gauge symmetries
that leave the action describing the theory unchanged.
As a result, both of these local spacetime symmetries lead to Noether identities.

First, consider diffeomorphism invariance using a metric description.
A generic form of the action in GR consists of an Einstein-Hilbert term
and a matter term,
\beq
S= \int d^4x \sqrt{-g} \left[ \fr 1 2 R + {\cal L}(g_{\mu\nu}, f^\ps) \right] .
\label{Smetric}
\eeq
Here, units with $\hbar = c = 8 \pi G = 1$ are used.
The matter term ${\cal L}$ depends on the metric tensor, $g_{\mu\nu}$, and 
dynamical matter fields generically represented as tensors $f^\ps$,
where $\ps$ collectively denotes the relevant tensor indices.
For simplicity, it is assumed that ${\cal L}$ depends on $f^\ps$ and at most 
first covariant derivatives of $f^\ps$.

Under local diffeomorphisms, with vector parameters, $\xi^\mu$,
tensor fields transform with changes given as Lie derivatives,
while the spacetime coordinates remain unchanged.
Specifically, $f^\ps \rightarrow f^\ps + {\cal L}_\xi f^\ps$, and
\beq
g_{\mu\nu} \rightarrow g_{\mu\nu} + {\cal L}_\xi g_{\mu\nu}
= g_{\mu\nu} + D_\mu \xi_\nu + D_\nu \xi_\mu .
\label{gLie}
\eeq
In GR, $S$ is invariant under these transformations.

To reveal the LLI, 
a vierbein description is used,
where the metric is defined in terms of the vierbein as
\beq
g_{\mu\nu} = \vb \mu a \vb \nu b \et_{ab} .
\label{gee}
\eeq
Here, Latin indices denote components with respect to a local Lorentz basis
in the tangent plane at each spacetime point,
while Greek indices are reserved for spacetime components.

For simplicity in this investigation,
a vierbein formalism restricted to Riemann spacetime (with no torsion)
is used, and spinor fields are not included.
Therefore, the resulting Einstein equations have the same structure as in GR,
except that contributions to $T^{\mu\nu}$ can involve local
matter fields as well as the vierbein,
which both transform under local Lorentz transformations.
It is in this way that LLI becomes a relevant local symmetry in GR.

Using such a vierbein approach, the action in GR becomes
\beq
S= \int d^4x \, e \left[ \fr 1 2 R + {\cal L}(\vb \mu a, f^\ps_{\pt{\ps}y}) \right] ,
\label{Svier}
\eeq
where $e$ is the determinant of the vierbein.
The tensors denoted here as $f^\ps_{\pt{\ps}y}$ can carry both 
spacetime and local indices,
where $y$ collectively denotes the local Lorentz indices.

Local Lorentz transformations are made with
respect to a given local basis. 
An infinitesimal transformation of $f^\ps_{\pt{\ps}y}$ has the form
\beq
 f^\ps_{\pt{\ps}y} \rightarrow  f^\ps_{\pt{\ps}y} + \fr 1 2 \ep^{ab} (X_{[ab]})_{\pt{y}y}^{x} f^\ps_{\pt{\ps}x} ,
 \label{LLT}
 \eeq
where $\ep_{ab} = - \ep_{ba}$ are the six infinitesimal 
parameters for the local Lorentz group, 
and $(X_{[ab]})_{\pt{y}y}^{x}$
gives an irreducible representation.
As a specific example,
the vierbein, $\vb \mu a$, transforms as a contravariant vector
under infinitesimal local Lorentz transformations.
Putting in the appropriate representation,
the transformation takes the form 
\beq
\vb \mu a \rightarrow \vb \mu a + \ep^a_{\pt{a}b} \vb \mu b .
\label{LTe}
\eeq

At the same time, 
$\vb \mu a$ transforms as a covariant spacetime vector under diffeomorphisms,
and $f^\ps_{\pt{\ps}y}$ transforms as a spacetime tensor with indices denoted by $\ps$.
The action in \rf{Svier} is then invariant under both 
diffeomorphisms and local Lorentz transformations.

\subsection{Observer transformations}

The spacetime coordinate system and local Lorentz basis do
not change under diffeomorphism and local Lorentz gauge transformations.
Only the physical gravitational and matter fields transform.
At the same time, the action in GR is mathematically
invariant under both general coordinate transformations
and rotations of the Lorentz bases in local frames.
These mathematical invariances of the action are referred to here, respectively, 
as general coordinate invariance (GCI) and observer LLI.

In GR,
it can be shown that infinitesimal general coordinate transformations,
$x^\mu \rightarrow {x^\prime}^\mu = x^\mu - \xi^\mu$,
combined with an expansion of the Lagrangian in $\xi^\mu$ 
and relabeling of coordinates, 
result in transformations of the fields that have the same
mathematical form as the diffeomorphism gauge transformations.
Likewise, in a vierbein treatment,
if inverse parameters, $-\ep_{ab}$, are used to rotate the local Lorentz basis,
the resulting infinitesimal transformations of local tensor components are
mathematically the same as the local Lorentz transformations.

In GR, 
and in the absence of symmetry breaking,
these can be viewed as active and passive transformations,
which are inversely related.
The gauge symmetries in fixed coordinate and local frames
are the active transformations,
while GCI and observer LLI are passive transformations
that move between different observer frames. 

\subsection{Noether Identities in GR}

Historically, Noether proved two theorems that are important in field theory
\cite{en18}.
The first involves global symmetry groups, 
with transformation parameters that are constant.
It states that a conserved current can be associated with invariance
of the action under a global symmetry.
The second theorem involves local symmetry groups,
with parameters that are spacetime dependent.
It results in identities that must hold when a theory is invariant under a local
symmetry group.

In GR, 
diffeomorphisms and LLI are local symmetries,
and therefore according to the second theorem
there are associated Noether identities
\cite{A62,Traut62,bb2002,Obuk}.
However, in GR, the same mathematical transformations in the action 
can be obtained using either the active forms 
due to diffeomorphism invariance and LLI or using the 
passive transformations associated with GCI and observer LLI
\cite{W72}. 
Thus, the same Noether identities can be obtained using either type 
of transformation in the context of GR.

At the same time,
it is important to keep in mind that there are differences
between active and passive spacetime transformations.
For example,
the active transformations are symmetry transformations,
where invariance of the action only holds when the symmetries are unbroken.
It is for this reason that nondynamical background fields are not permitted in GR.
On the other hand,
any realistic action is required to be invariant under the observer 
transformations so that observer independence is maintained.
In this case, invariance is imposed on the action regardless of 
whether there are background fields or not.
Nonetheless, since GR does not allow fixed background fields,
the Noether identities resulting from GCI and observer LLI 
are mathematically the same in the context of GR as those due to 
diffeomorphism invariance and LLI in GR.

The Noether identity stemming from diffeomorphism invariance in GR
(or equivalently GCI) is obtained by performing field variations in the action 
given in \rf{Smetric},
where the variations in the metric and matter fields are given as Lie derivatives, 
with local parameters $\xi^\mu$, 
acting on these fields.
The invariance of the action gives 
\bea
\int d^4x \, \left[ \fr 1 2 \fr {\de (\sqrt{-g} R)} {\de g_{\mu\nu}} {\cal L}_\xi g_{\mu\nu} 
\right.
\quad\quad\quad\quad\quad\quad\quad\quad\quad
\nonumber \\
\left.
+ \fr {\de (\sqrt{-g} {\cal L})} {\de g_{\mu\nu}} {\cal L}_\xi g_{\mu\nu}
+ \sqrt{-g} \fr {\de {\cal L}} {\de f^\ps}  {\cal L}_\xi f^\ps \right] = 0 .
\label{diffvarL}
\eea
Putting in the expressions for the Lie derivatives,
which depend on the type of the tensor $f^\ps$, 
and performing integrations by parts, 
gives a generic expression 
\bea
\int d^4x \, \sqrt{-g} \, \xi_\nu [ D_\mu ( G^{\mu\nu} - T^{\mu\nu} ) 
\quad\quad\quad\quad\quad\quad\quad\quad
\nonumber
\\
+ \fr {\de {\cal L}} {\de f^\ps} \ga^{\ps\nu} + D_\mu (\fr {\de {\cal L}} {\de f^\ps} \ga^{\ps\mu\nu} ) ] = 0 .
\label{NoetherGRintegral}
\eea
Here, the spacetime indices represented by $\ps$ are summed,
and the coefficients $\ga^{\ps\nu}$ and $\ga^{\ps\mu\nu}$ 
represent general functions of the field variables,
which depend on the form of $f^\ps$.

The Noether identity associated with diffeomorphism invariance
follows from requiring that the intergral 
in \rf{NoetherGRintegral} must vanish for all parameters $\xi_\nu$
that vanish on the boundary surfaces.
This results in four identities that
must be obeyed by the metric and matter fields,
which have the generic form
\beq
D_\mu ( G^{\mu\nu} - T^{\mu\nu} ) 
+ \fr {\de {\cal L}} {\de f^\ps} \ga^{\ps\nu} + D_\mu (\fr {\de {\cal L}} {\de f^\ps} \ga^{\ps\mu\nu} ) = 0 ,
\label{NoetherGR}
\eeq
The first term in these identities contains the Euler-Lagrange expression for the metric,
$(G^{\mu\nu} - T^{\mu\nu})$,
while $\fr {\de {\cal L}} {\de f^\ps}$ represents the Euler-Lagrange
expression for the tensor $f^\ps$.
For example, if $f^\ps$ is a contravariant vector with index $\ps$ replaced by $\al$, 
and ${\cal L}$ depends on both $f^\al$ and its first derivative, then
\beq
\fr {\de {\cal L}} {\de f^\al} \equiv - D_\mu \left( \fr {\partial {\cal L}} {\partial D_\mu f^\al} \right) + \fr {\partial {\cal L}} {\partial f^\al} .
\label{EL}
\eeq
Note that as an identity, \rf{NoetherGR} holds off shell,
i.e., the Euler-Lagrange expressions need not vanish.

The usual interpretation of this Noether identity in GR 
is that four of the field components in the theory are not
dynamically independent from the others.
Thus, if all but four Euler-Lagrange equations are set to zero,
then the remaining four equations of motion must hold automatically.
In particular, by combining \rf{NoetherGR} with the contracted Bianchi identity,
$D_\mu G^{\mu\nu} = 0$,
it also follows that $D_\mu T^{\mu\nu} = 0$ automatically holds in GR
when the matter fields are on shell,
obeying $\fr {\de {\cal L}} {\de f^\ps} = 0$.

In GR, the loss of four independent degrees of freedom is expected 
as a result of the local gauge invariance under diffeomorphisms,
where four field components can be fixed or set to zero by 
imposing gauge-fixing conditions.
Such gauge-fixing conditions, or equivalently coordinate-fixing conditions,
are in fact needed in order to be able to carry 
out and solve the initial-value problem in GR.
Only with these extra conditions can an unambiguous evolution
be determined for the dynamical degrees of freedom, 
subject to their initial constraints.

To examine Noether's identity associated with LLI,
a vierbein approach can be used.
The action in the absence of spin and torsion
then has the form given in \rf{Svier},
and fields with local tensor indices transform,
as in \rf{LLT} for example,
under local Lorentz transformations.  
In this case, the corresponding Noether identity is obtained by performing variations having
the form of local Lorentz transformations.
The resulting identity consists of six equations,
\bea
(G^{\mu\nu} - T^{\mu\nu}) (\lvb \mu a \lvb \nu b - \lvb \mu b \lvb \nu a)
\quad\quad\quad\quad\quad\quad\quad
\nonumber
\\
+ \fr 1 2 \fr {\de {\cal L}} {\de f^\ps_{\pt{\ps}y}}  (X_{[ab]})_{\pt{y}y}^{x} f^\ps_{\pt{\ps}x} = 0 ,
\label{LLINoether}
\eea
which hold off shell.
When the matter fields are on shell, obeying $ \fr {\de {\cal L}} {\de f^\ps_{\pt{\ps}y}} = 0$,
multiplying by inverse vierbeins and using
the symmetry of the Einstein tensor
shows that it reduces to the requirement that the energy-momentum 
tensor must be symmetric, obeying
$T^{\mu\nu} = T^{\nu\mu}$.

\section{Nondynamical Backgrounds}

When a field theory includes a nondynamical background field,
the usual notion of an active symmetry transformation no longer applies.
This is because the background field is understood to be unchangeable,
and it therefore cannot be transformed.  
In this context,
it is useful and more common to distinguish what are known 
as particle and observer transformations
\cite{akgrav04}.
These are the types of transformations that are relevant 
when there is spacetime symmetry breaking.

Particle transformations are the same as active transformations 
when they act on any fields other than the background fields.
However, when a particle transformation acts on a background field
the background remains fixed and does not transform.
Particle transformations therefore correspond to the physical transformations 
that can be performed in a laboratory,
where it is not possible for an experimenter to alter the background fields.
Notice that particle transformations keep the coordinate system and local Lorentz basis unchanged.
In contrast,
observer transformations are passive changes of the coordinates and local bases.
Under observer transformations, all tensor components transform,
including those of the background fields.

Observer independence requires that the Lagrangian must be a 
scalar under observer spacetime transformations.
Thus, when couplings to a fixed background appear in a theory,
it is the particle symmetries that are broken either spontaneously or explicitly,
while GCI and observer LLI remain mathematical invariances of the action.

If the background field is dynamical,
forming as a result of spontaneous breaking of the particle symmetries,
then it satisfies the equations of motion as a vacuum solution.
In this case, the interpretation of Noether's identities is the same as in GR.
This is true as well when excitations about the background occur,
since these include Nambu-Goldstone and massive modes 
that restore the spontaneously broken particle symmetries.

However, when a background field is nondynamical,
particle diffeomorphisms and LLI are broken explicitly,
and the Euler-Lagrange expressions for the background fields need not vanish.
Therefore, the usual interpretation of the Noether identities does not apply,
and instead a modified interpretation is needed.

\subsection{Explicit Symmetry Breaking}

To investigate Noether's identities with explicit diffeomorphism breaking,
consider a gravitational theory with a fixed nondynamical background,
which is denoted generically as a tensor $\bar k_{\la\mu\nu\cdots}$.
In a metric description,
the action is
\beq
S= \int d^4x \sqrt{-g} \left[ \fr 1 2 R + {\cal L}(g_{\mu\nu}, f^\ps, \bar k_{\la\mu\nu\cdots}) \right] ,
\label{brokenSmetric}
\eeq
where ${\cal L}$ is an observer scalar that depends on the dynamical matter fields, $f^\ps$, 
and their first derivatives, as well as on the nondynamical background, $\bar k_{\la\mu\nu\cdots}$.
With this form, the action $S$ explicitly breaks particle diffeomorphisms
while maintaining observer GCI. 

To consider local Lorentz transformations,
the theory can be redefined using dynamical vierbeins, $\vb \mu a$, in place of the metric.
At the same time, the background field can be written in terms of tensor components given
with respect to both the spacetime frame and the local Lorentz frame,
where in the latter case they are denoted as $\bar k_{abc\cdots}$.
While the dynamical vierbein and matter fields transform under 
particle diffeomorphisms and local Lorentz transformations,
the nondynamical background tensor, $\bar k_{\la\mu\nu\cdots}$, remains fixed.
Since the coordinate system and local basis do not change either
under these particle transformations,
the components of the background defined with respect to the local Lorentz basis,
$\bar k_{abc \cdots}$, must remain fixed as well.
This means that these different fixed components must be related by a
nondynamical vierbein, denoted as $\barvb \mu a$,
which is also fixed under particle diffeomorphisms and local Lorentz transformations.
The defining relation for the background vierbein is
\beq
\bar k_{\la\mu\nu\cdots} = \barvb \la a \barvb \mu b \barvb \nu c \cdots \bar k_{abc \cdots} ,
\label{ebar}
\eeq
where every quantity in this expression remains fixed under particle
diffeomorphisms and local Lorentz transformations.

The action replacing \rf{brokenSmetric} can then be written as
\beq
S= \int d^4x e \left[ \fr 1 2 R + {\cal L}(\vb \mu a,\barvb \mu a,f^\ps_{\pt{\ps}y}, \bar k_{abc\cdots}) \right] ,
\label{brokenSvier}
\eeq
where ${\cal L}$ now depends on both the physical vierbein, $\vb \mu a$,
and the background vierbein, $\barvb \mu a$,
in addition to the dynamical matter fields and the nondynamical background.
This form of the action explicitly breaks diffeomorphisms and local Lorentz transformations,
but it is mathematically invariant under observer general coordinate transformations and
changes of local Lorentz bases,
since all of the field components in ${\cal L}$ transform under
the observer transformations.

\subsection{Noether Identities with Fixed Backgrounds}

The Noether identities that hold when fixed backgrounds are present 
stem from the observer invariances in the action.
The identities are obtained by performing mathematical field variations in the 
action having the form of the local observer transformations.
As in GR,
both a metric formalism and a vierbein formalism can be considered.

\subsubsection{Metric Formalism}

In a metric formalism, the action is given in \rf{brokenSmetric},
and the Noether identity associated with observer
GCI is found to have the general form
\bea
D_\mu ( G^{\mu\nu} - T^{\mu\nu} ) 
+ \fr {\de {\cal L}} {\de f^\ps} \ga^{\ps\nu} + D_\mu (\fr {\de {\cal L}} {\de f^\ps} \ga^{\ps\mu\nu} ) 
\quad\quad\quad\quad
\nonumber \\
+ \fr {\de {\cal L}} {\de \bar k_{\al\be\ga\cdots}} \la^{\nu}_{\al\be\ga\cdots} 
+ D_\mu (\fr {\de {\cal L}} {\de \bar k_{\al\be\ga\cdots}}  \la^{\mu\nu}_{\al\be\ga\cdots} ) 
= 0 .
\quad
\label{obsNoetherGR}
\eea
Here, the coefficients $\ga^{\ps\nu}$, $\ga^{\ps\mu\nu}$, 
$\la^{\nu}_{\al\be\ga\cdots}$ and $ \la^{\mu\nu}_{\al\be\ga\cdots}$
denote functions of the field components,
where their specific forms depend on how ${\cal L}$ is defined.
Note that these are off-shell equations that hold for all values of the fields.

The potential conflict that arises when diffeomorphisms are explicitly broken
becomes evident when the dynamical matter fields are put on shell.
Setting $ \fr {\de {\cal L}} {\de f^\ps} = 0$ and using the contracted Bianchi
identity, $D_\mu G^{\mu\nu} = 0$, reveals that $D_\mu T^{\mu\nu}$ 
can only vanish when the remaining two terms in the Noether identity vanish.
At the same time, consistency with Einstein's equations requires that
$D_\mu T^{\mu\nu} = 0$ must hold on shell.
Therefore theoretical consistency requires that on shell the following must hold:
\beq
 \fr {\de {\cal L}} {\de \bar k_{\al\be\ga\cdots}} \la^{\nu}_{\al\be\ga\cdots} 
+ D_\mu (\fr {\de {\cal L}} {\de \bar k_{\al\be\ga\cdots}}  \la^{\mu\nu}_{\al\be\ga\cdots} ) 
= 0 .
\label{diffconstraint}
\eeq
However, unlike the matter fields,
the nondynamical background fields need not have
vanishing Euler-Lagrange equations,
which allows
\beq
\fr {\de {\cal L}} {\de \bar k_{\al\be\ga\cdots}} \ne 0 .
\label{nondyn}
\eeq
Thus, \rf{diffconstraint},
which contains the fixed background, 
becomes a consistency condition that must be satisfied
by the dynamical fields in the theory.

In general it is possible for the four equations in \rf{diffconstraint} to have solutions
because the loss of diffeomorphism invariance
means that there are four additional degrees of freedom
in comparison to GR.
These are the degrees of freedom that in a gauge-invariant theory
such as GR would normally be gauged away.
However, explicit breaking no longer allows this.
In a metric formalism,
it is natural to let the metric tensor have the four additional degrees of freedom.
In that case,
\rf{diffconstraint} can be viewed as four constraints that can be
satisfied by the metric due to its having four additional degrees of freedom.

At the same time,
inconsistency can arise when \rf{diffconstraint} is found not to hold.
For example,
there must be sufficient coupling with the metric so that at least
four components appear in these equations.
If a specific ansatz for the metric is assumed,
as in a spatially homogeneous and isotropic universe,
there might not be enough degrees of freedom in the metric
to satisfy \rf{diffconstraint}.
This can lead to certain geometries being ruled out
in the presence of incompatible backgrounds.
In other cases, if there are not enough metric components
but matter couplings occur,
this can put additional conditions on the matter fields,
which in turn can lead to consistency issues in the matter sector.

Note that even if consistency is maintained,
the interpretation of Noether's identity when there is explicit breaking
is fundamentally different from the interpretation in GR.
For example,
in GR the metric affects the curvature,
which remains largely distinguishable from matter dynamical 
effects even when both have backreactions on the other. 
However, with explicit breaking,
the metric affects not only the curvature but it must
also absorb the physical backreactions that the fixed
background tensor is unable to have.

It is also possible to interpret Noether's identity when a background
field is present as providing a set of 
coordinate-fixing conditions for the metric.
The specific form of the conditions depends on ${\cal L}$ and
the values given for $\bar k_{\la\mu\nu\cdots}$ in a particular coordinate frame.
Normally GCI allows freedom in the choice of coordinates.
However, with explicit breaking,
when the components of a fixed background tensor are
given specified values, $\bar k_{\la\mu\nu\cdots}$, 
this must be done with respect to a particular coordinate frame.
It is then the conditions from the Noether identity that makes the
chosen coordinate frame compatible with the dynamical equations of motion.
This in turn allows covariant energy-momentum conservation to hold.

As an illustrative example, if explicit breaking is caused by
a term with ${\cal L} = g^{\mu\nu} \et_{\mu\nu}$ in the action,
where $\et_{\mu\nu} = {\rm diag}(-1,1,1,1)$ in the chosen coordinate frame,
then the conditions in \rf{diffconstraint} can be shown to reduce to
the requirement that $g^{\mu\nu} \Ga^\la_{\mu\nu}=0$.
In GR, these are the conditions for harmonic coordinates,
which are commonly chosen as gauge-fixing conditions. 
The difference here is that they emerge as a result of 
the Noether identity associated with GCI
and the choice of ${\cal L}$ and the 
background field $\et_{\mu\nu}$.

Theories with explicit diffeomorphism breaking differ from
GR in other important respects as well.
For example, theories with nondynamical backgrounds
have a different constraint structure, 
and therefore the number of physical degrees of freedom can differ
\cite{rbngrpav}.
As a result,
the propagation of gravitational excitations is typically modified.
Based on these effects,
models can be investigated and in some cases ruled out if they permit 
unphysical degrees of freedom (such as ghost modes) or if they contradict experiments.
While these are serious issues concerning the physical viability of theories
with nondynamical backgrounds,
they are not explored further here.
Instead,
the focus here remains on the Noether identities,
which in many respects are more restrictive than phenomenological tests,
since the Noether identities provide consistency conditions that 
must hold purely on theoretical grounds.

\subsubsection{Vierbein Formalism}

Using a vierbein formalism,
with the action given in \rf{brokenSvier},
Noether identities associated with both GCI
and observer LLI can be found.

The identity stemming from GCI
has a structure similar to \rf{obsNoetherGR} except that
the matter and background vierbein also carry local indices.
For the matter fields these are labeled generically using $y$
to denote the set of local indices.
As a result, the coefficients in the terms involving the
matter fields, $f^\ps_{\pt{\ps}y}$, carry local indices as well,
and these are written as $\ga^{\ps\nu}_{\pt{\ps\nu}y}$ and
$\ga^{\ps\mu\nu}_{\pt{\ps\mu\nu}y}$.
The background vierbein, $ \barvb \mu a$, and
local tensor, $\bar k_{abc\cdots}$, are respectively vectors
and scalars under general coordinate transformations,
and their contributions can be written explicitly.
The result is
\bea
D_\mu ( G^{\mu\nu} - T^{\mu\nu} ) 
+ \fr {\de {\cal L}} {\de f^\ps_{\pt{\ps}y}} \ga^{\ps\nu}_{\pt{\ps\nu}y} 
+ D_\mu (\fr {\de {\cal L}} {\de f^\ps_{\pt{\ps}y}} \ga^{\ps\mu\nu}_{\pt{\ps\mu\nu}y} ) 
\nonumber \\
-D_\mu \left( \fr {\de {\cal L}} {\de \barvb \mu a} g^{\nu\al} \barvb \al a \right)
+ \fr {\de {\cal L}} {\de \barvb \mu a} D^\nu \barvb \mu a
\quad\quad\quad\quad
\nonumber \\
\quad\quad\quad\quad\quad
+ \fr {\de {\cal L}} {\de \bar k_{abc\cdots}} D^\nu \bar k_{abc\cdots} 
= 0 .
\quad\quad\quad\quad\quad
\label{vierobsNoetherGR}
\eea

When the matter fields, $f^\ps_{\pt{\ps}y}$, are put on shell,
and the contracted Bianchi identity is used,
this identity shows that the last three terms involving
$ \barvb \mu a$ and $\bar k_{abc\cdots}$ must vanish
in order for covariant energy-momentum conservation to hold.
Since in this case, the Euler-Lagrange expressions for the
backgrounds need not vanish,
\beq
\fr {\de {\cal L}} {\de \barvb \mu a} \ne 0 ,
\quad\quad
\fr {\de {\cal L}} {\de \bar k_{abc\cdots}} \ne 0 ,
\label{novanishvier}
\eeq
there is again the possibility of conflicts and inconsistency.
To avoid these issues it must be that the physical vierbein,
which has four additional modes due to the lack of diffeomorphism invariance,
must take values that satisfy these conditions.
The question of whether this happens consistently or not then largely depends on
the extent to which vierbein couplings are included in the terms appearing in \rf{vierobsNoetherGR}.

The Noether identity stemming from observer LLI can be found as well.
It is obtained by performing infinitesimal observer Lorentz transformations in the action, 
where a general representation of the local Lorentz group is used for $f^\ps_{\pt{\ps}x}$,
while vector and tensor representations are used,
respectively, for $\barvb \mu a$ and $\bar k_{abc\cdots}$.
The result is
%\begin{widetext}
\bea
(G^{\mu\nu} - T^{\mu\nu}) (\lvb \mu a \lvb \nu b - \lvb \mu b \lvb \nu a)
\quad\quad\quad\quad\quad\quad\quad\quad\quad
\nonumber
\\
+ \fr 1 2 \fr {\de {\cal L}} {\de f^\ps_{\pt{\ps}y}}  (X_{[ab]})_{\pt{y}y}^{x} f^\ps_{\pt{\ps}x} 
+ \left( \fr {\de {\cal L}} {\de \barvb \mu a} \barlvb \mu b - \fr {\de {\cal L}} {\de \barvb \mu b} \barlvb \mu a \right)
\quad\quad
\nonumber
\\
+ \fr {\de {\cal L}} {\de \bar k_{cde\cdots}} \left[ (\et_{ac} \bar k_{bde\cdots} - \et_{bc} \bar k_{ade\cdots}) \right.
\quad\quad\quad\quad\quad\quad\quad\quad
\nonumber
\\
+ \left. (\et_{ad} \bar k_{cbe\cdots} - \et_{bd} \bar k_{cae\cdots}) \right.
\quad\quad\quad\quad\quad\quad\quad
\nonumber
\\
+  \left. (\et_{ac} \bar k_{cdb\cdots} - \et_{bc} \bar k_{cda\cdots})
+ \cdots \right] = 0 .
\quad
\label{vierLLINoether}
\eea
%\end{widetext}

When the matter fields are put on shell,
obeying $\fr {\de {\cal L}} {\de f^\ps_{\pt{\ps}y}}  = 0$,
and the symmetry of the Einstein tensor is used,
it follows that $T^{\mu\nu}$ is symmetric only if the
additional terms in \rf{vierLLINoether} either cancel or vanish.
Since Einstein's equations (in the absence of spin and torsion)
give the result that $T^{\mu\nu}$ must be symmetric,
theoretical consistency requires that
the additional terms associated with $\barvb \mu a$ and $\bar k_{abc\cdots}$
in \rf{vierLLINoether} must equal zero.
Since $\barvb \mu a$ and $\bar k_{abc\cdots}$ are nondynamical backgrounds,
their Euler-Lagrange expressions need not vanish,
as indicated in \rf{novanishvier}.
Therefore, the six conditions resulting from \rf{vierLLINoether} 
must be satisfied by the physical vierbein.
In general, this is possible since with violation of LLI
the vierbein has six additional degrees of freedom that would otherwise be gauge degrees of freedom.

In many cases cancellations can occur in \rf{vierLLINoether} that
reduce the Noether identity for observer LLI to a trivial identity.
This is the case when only locally Lorentz-invariant combinations of fields 
under particle Lorentz transformations appear in the Lagrangian.
For example if the dynamical vierbein only occurs in the combination
$g_{\mu\nu}=\vb \mu a \vb \nu b \et_{ab}$,
and the local backgrounds only appear through
$\bar k_{\la\mu\nu\cdots} = \barvb \la a \barvb \mu b \barvb \nu c \cdots \bar k_{abc \cdots} $,
then each of these combinations is invariant under particle local Lorentz transformations,
and the Noether identity reduces trivially to $T^{\mu\nu} = T^{\nu\mu}$.
It is only when there is mixing between the dynamical and nondynamical local tensors
in combinations that are not invariant under particle local Lorentz 
transformations that a nontrivial identity results.

As an illustrative example, if explicit breaking of LLI 
is caused by a term with ${\cal L} = \ivb \mu a \barvb \mu a$ in the action, 
and the matter fields are put on shell,
then \rf{vierLLINoether} combined with requiring $T^{\mu\nu} = T^{\nu\mu}$ 
reduces to a set of symmetry conditions for the physical and background vierbeins,
\beq
\ivb \mu a \barlvb \mu b - \ivb \mu b \barlvb \mu a = 0 .
\label{syme}
\eeq
These six constraints,
which depend on the specific values of the background vierbein components,
must be satisfied by the physical vierbein.

For example, if the physical and background vierbeins are expanded in 
infinitesimal linear approximations about flat backgrounds,
\bea
\ivb \mu a \simeq \de^\mu_{\pt{\mu}a} - \half h^\mu_{\pt{\mu}a} + \ch^\mu_{\pt{\mu}a} ,
\nonumber \\
\barlvb \mu a \simeq \et_{\mu a} + \half \bar h_{\mu a} + \bar \ch_{\mu a} ,
\label{infexpans}
\eea
where $h^\mu_{\pt{\mu}a}$ and $\bar h_{\mu a}$ are symmetric components,
and $\ch^\mu_{\pt{\mu}a}$ and $ \bar \ch_{\mu a}$ are antisymmetric components,
then the conditions in \rf{syme} reduce to 
\beq
\ch_{ab} \simeq \bar \ch_{ab} .
\label{antich}
\eeq
Consistency in this case requires that the
six antisymmetric components of the dynamical vierbein must
equal the six anti-symmetric components of the background vierbein.

With explicit violation of LLI,
it is possible to interpret the Noether identity stemming from observer LLI
as providing coordinate-fixing conditions for the vierbein,
such as \rf{syme}. 
This is similar to how the conditions 
stemming from observer GCI can be interpreted as 
coordinate-fixing conditions for the metric.
Indeed, the conditions in \rf{syme} have been used 
as gauge-fixing conditions in quantum-gravity 
calculations where there is no Lorentz violation,
but where a perturbative background-field formalism is used.
In that context,
the conditions \rf{syme} fix the vierbein
to what is called the Deser-van Nieuwenhuizen gauge
\cite{dvn74}.
Evidently,
as the above example shows,
these same conditions can arise when there is explicit Lorentz breaking.
However, with Lorentz violation,
they are not a gauge choice.
Instead, they are conditions imposed by the Noether identity.

\section{Applications}

In this section,
specific models with nondynamical background fields are examined.
This permits a more detailed look at how the Noether identities impose
conditions that either allow a theory to be consistent or rule it out.
Models with spacetime fields and backgrounds must obey the identity
in \rf{obsNoetherGR},
while models in a vierbein formalism must obey both identities
\rf{vierobsNoetherGR} and \rf{vierLLINoether}.

\subsection{Scalar and Tensor Backgrounds}

To begin,
consider gravitational models described using a metric formalism.
In this case the Noether identity is given in \rf{obsNoetherGR},
and as previously noted the consistency requirements that follow 
depend in large part on the type of background field that is present
and how it couples to the metric tensor.

The most restrictive cases involve nondynamical scalar backgrounds.
Because scalars do not couple directly to the metric,
there are greater limitations on how the additional metric modes 
(which would otherwise be gauge degrees of freedom)
can appear in the consistency conditions stemming from Noether's identity.

To examine this further,
consider a set of nondynamical scalars, $\bar \ph^A$,
labeled with an internal index $A$.
The on-shell consistency conditions stemming from  \rf{obsNoetherGR} then have the form
\beq
\fr {\de {\cal L}} {\de \bar \ph^A} \, \partial_\nu \bar \ph^A = 0,
\label{phiA}
\eeq
where the index $A$ is summed.
It is assumed here that $\partial_\nu \bar \ph^A \ne 0$ so that
all four diffeomorphisms are explicitly broken by the background scalars.
Evidently, when this is the case,
and the derivatives are functionally independent,
then the Euler-Lagrange equations,
$\fr {\de {\cal L}} {\de \bar \ph^A} = 0$,
must hold for the scalars despite the fact that they are nondynamical.

An example involving a single background scalar is Chern-Simons gravity in four spacetime dimensions.
The Lagrangian $\sqrt{-g} {\cal L} \sim \th \, ^\ast R R $,
where $\th$ is a nondynamical scalar and $^\ast R R $ is the gravitational Pontryagin density
\cite{rjsp,rb15a}.
In this case, the Euler-Lagrange equation for the scalar gives the condition
\beq
\fr {\de {\cal L}} {\de \th} \sim \, {^\ast R} R = 0 ,
\label{thetaeq}
\eeq
which must be satisfied by the metric.
Thus, the spacetime must either have a vanishing Pontryagin density
or the theory is inconsistent.

In models with Lagrangian terms 
${\cal L} (g_{\mu\nu}, \bar \ph^A, \partial_\mu \bar \ph^A)$,
which are functions of the metric, the scalars, 
and first derivatives of the scalars with minimal couplings,
the consistency conditions become
\beq
\left[ -D_\mu \fr {\partial {\cal L}} {\partial \partial_\mu \bar \ph^A}  
+ \fr {\partial {\cal L}} {\partial \bar \ph^A}\right] \partial_\nu \bar \ph^A = 0.
\label{phiA2}
\eeq
Since the scalars are nondynamical,
it is the extra degrees of freedom in the metric that must satisfy these conditions.
The metric in this case appears in the covariant derivatives, $D_\mu$,
which depend on the connection, $\Ga^\la_{\mu\nu}$.

The conditions in \rf{phiA2} require that four combinations of the 
Euler-Lagrange expressions must vanish.
However,
in theories with four background scalars, $\bar \ph^A$,
which explicitly break all four diffeomorphisms,
then the derivatives $\partial_\nu \bar \ph^A$ will in general
be functionally independent.
In this case, 
the solutions for the metric that satisfy \rf{phiA2}
must also satisfy the Euler-Lagrange equations for $\bar \ph^A$
by causing the terms in brackets to vanish.
This gives the appearance that the scalars $\bar \ph^A$ are dynamical,
since their Euler-Lagrange equations hold,
but in fact they are not.

On the other hand, if potential terms ${\cal L} (g_{\mu\nu},\bar k_{\la\mu\nu \cdots})$
are formed using only the metric, a background tensor $\bar k_{\la\mu\nu \cdots}$,
and no derivatives,
then the consistency conditions in this case are the four equations
\bea
- D_\la \left( \fr {\partial {\cal L}} {\partial \bar k_{\la\mu\nu \cdots}} \bar k_{\al\mu\nu \cdots} \right) 
- D_\mu \left( \fr {\partial {\cal L}} {\partial  \bar k_{\la\mu\nu \cdots}} \bar k_{\la\al\nu \cdots} \right) - \cdots
\nonumber \\
+  \fr {\partial {\cal L}} {\partial \bar k_{\la\mu\nu \cdots}} D_\al \bar k_{\la\mu\nu \cdots} = 0 .
\quad\quad
\label{kmunudots}
\eea
In this case, with tensor backgrounds,
the Euler-Lagrange expressions for the background tensors do not need to vanish,
and $\fr {\partial {\cal L}} {\partial \bar k_{\la\mu\nu \cdots}} \ne 0$ can hold.

Notice that regardless of how many components a nondynamical background,
e.g., $\bar k_{\la\mu\nu \cdots}$, might have,
there are always four consistency conditions that follow from 
the Noether identity associated with GCI.
At the same time,
specifying definite values for the components $\bar k_{\la\mu\nu \cdots}$
depends on the choice of a four-dimensional  coordinate system.

It is possible, however, to consider a form of the background tensor that
does not depend specifically on the choice of coordinate system.
For example, in a different coordinate system, 
the components of the tensor background 
can be obtained from the original components by 
making a general coordinate transformation.
The new coordinates become functions of the original coordinates,
and they can therefore be written as four independent scalar functions, 
$\bar \ph^A (x)$, with labels $A=0,1,2,3$.
Note that $\bar \ph^A$ are nondynamical scalars,
since both the original and transformed coordinate systems are specified.
The relationship between the original and transformed background tensors 
can then be written as
\beq
\bar k_{\la\mu\nu \cdots} (x) = \partial_\la \bar \ph^A \partial_\mu \bar \ph^B 
\partial_\nu \bar \ph^C \cdots \bar k_{ABC\cdots} (\bar \ph) .
\label{ktransf}
\eeq
Here,
$\bar k_{ABC\cdots} (\bar \ph)$ are scalar functions
that have the same functional dependence as $\bar k_{\la\mu\nu \cdots} (x)$.
Notice that if $\bar \ph^A = \de^A_\mu x^\mu$,
then the new and original coordinates are the same,
and $\bar k_{ABC\cdots} (\bar \ph)$ equals $\bar k_{\la\mu\nu \cdots} (x)$.

If this expression for $\bar k_{\la\mu\nu \cdots}$ is substituted into the Lagrangian, 
then ${\cal L}$ can be considered as a function of the 
metric and the four nondynamical scalars, $\bar \ph^A$,
with functional dependence given as
$ {\cal L}(g_{\mu\nu}, \bar \ph^A, \partial_\mu \bar \ph^A)$.
It therefore changes from a model depending on an unspecified number of background components,
$\bar k_{\la\mu\nu \cdots}$, to one that depends only on four fixed scalars, $\bar \ph^A$, and their derivatives,
$\partial_\mu \bar \ph^A$.

In making the switch from the description with $\bar k_{\la\mu\nu \cdots}$ to one depending
on $\bar \ph^A$,
the conditions stemming from the Noether identities change as well.
Instead of the conditions in \rf{kmunudots},
where the Euler-Lagrange equations for $\bar k_{\la\mu\nu \cdots}$ need not hold,
the new conditions are those in \rf{phiA2}. 
In this case, the metric must provide solutions to the 
Euler-Lagrange equations for the scalars, $\bar \ph^A$,
which gives the appearance that the scalars are then dynamical fields.
However, the scalars $\bar \ph^A$ remain fixed nondynamical backgrounds
that are related to fixing the coordinates so that compatibility with GCI is maintained.
In both \rf{phiA2} and \rf{kmunudots}, 
it is the existence of the additional metric degrees of freedom that allows these equations to hold.

\subsection{St\"uckelberg Trick}

Since changing the background fields from $\bar k_{\la\mu\nu \cdots}$ to $\bar \ph^A$,
gives solutions of \rf{phiA2}
where the Euler-Lagrange equations for $\bar \ph^A$ are required to hold,
this makes possible a trick known as the St\"uckelberg trick
\cite{ags03}.
The trick is to let the four scalars, $\bar \ph^A$, be dynamical,
which restores diffeomorphism invariance.
While the St\"uckelberg approach introduces four additional dynamical degrees of freedom,
it also creates four gauge degrees of freedom by restoring diffeomorphism invariance.
Thus, the number of independent degrees of freedom does not change.

With scalars $\bar \ph^A$ that are now dynamical,
the usual interpretation of Noether's identity 
associated with diffeomorphism invariance in GR becomes applicable.
Notice, however, that with gauge invariance restored
the individual Euler-Lagrange equations,
\beq
-D_\mu \fr {\partial {\cal L}} {\partial \partial_\mu \bar \ph^A}  
+ \fr {\partial {\cal L}} {\partial \bar \ph^A}  = 0 ,
\label{phiA3}
\eeq
involve both the dynamical scalars and the metric.
In particular,
if the diffeomorphism gauge freedom is used to set the scalars equal to 
the coordinates, so that $\bar \ph^A = \de^A_\mu x^\mu$,
then the metric no longer has four gauge degrees of freedom.
In this case,
it is again extra degrees of freedom in the metric that must give solutions of \rf{phiA3},
which therefore parallels the theory with explicit breaking.

The St\"uckelberg trick is widely used in theories of massive gravity,
where it appears to eliminate the awkwardness 
associated with having nondynamical backgrounds and explicit diffeomorphism breaking.
However, it is important to realize that
the reason the trick works hinges on the fact
that when nondynamical scalars $\bar \ph^A$ are introduced 
in place of $\bar k_{\la\mu\nu \cdots}$
the consistency conditions following from Noether's identity require that 
the Euler-Lagrange equations for the background scalars must hold.
Without this requirement,
it would not be possible to let the scalars be dynamical and
still have an equivalent theory.  

Lastly, note that even with the St\"uckelberg trick 
the original backgrounds, $\bar k_{\la\mu\nu \cdots}$,
never satisfy their Euler-Lagrange equations
and do not become dynamical.
The backgrounds $\bar k_{\la\mu\nu \cdots}$ become somewhat hidden in
the St\"uckelberg approach, 
where they become dependent on the gauge fixing and choice of coordinates.
Only the four scalar degrees of freedom become dynamical using
the St\"uckelberg approach.

\subsection{Einstein-Maxwell with a Fixed Vector Current}

As an example of explicit breaking that contains a dynamical matter field in addition 
to the metric and the nondynamical background,
consider Einstein-Maxwell theory with a nondynamical external vector current.
The dynamical Maxwell vector field is given as a covariant vector $A_\mu$,
with field strength $F_{\mu\nu} = \partial_\mu A_\nu - \partial_\nu A_\mu$,
while the nondynamical background is a 
contravariant vector $\bar k^\mu$ that couples directly to $A_\mu$.
This gives the background the form of a charge current, $\bar k^\mu = J^\mu$.
However, the current $J^\mu$ in this case is not dynamical,
and it does not have backreactions.

The action in this case is
\beq
S= \int d^4x \sqrt{-g} \left[ \fr 1 2 R - \fr 1 4 F_{\mu\nu} F^{\mu\nu} + A_\mu \bar k^\mu \right] .
\label{EMJ}
\eeq
With these definitions there is no coupling with the metric
in the interaction term, $A_\mu \bar k^\mu$.
Diffeomorphism invariance and LLI are explicitly broken in $S$ due to the
presence of the nondynamical background, $\bar k^\mu$.
However, GCI still holds.
A Noether identity with the form given in \rf{obsNoetherGR} therefore follows,
where $A_\mu$ replaces $f^\ps$ and $\bar k^\mu$ is the background tensor.
In this context, the identity can be written as
\bea
g_{\nu\al} D_\mu (G^{\mu\nu} - T^{\mu\nu}) - D_\nu [(D_\mu F^{\mu\nu} + \bar k^\nu) A_\al]
\nonumber \\
+ (D_\mu F^{\mu\nu} + \bar k^\nu)  D_\al A_\nu 
\quad\quad\quad\quad\quad
\nonumber \\
+ D_\mu \left( \fr {\partial {\cal L}} {\partial \bar k^\al} \bar k^\mu \right)
+ \fr {\partial {\cal L}} {\partial \bar k^\mu} D_\al \bar k^\mu = 0 .
\label{EMNoether}
\eea

From this identity it follows that when the Einstein equations, $G^{\mu\nu} = T^{\mu\nu}$,
and Maxwell equations, $D_\mu F^{\mu\nu} = -\bar k^\nu$, both hold,
the sum of the last two terms in \rf{EMNoether} must therefore vanish.
However, $\bar k^\mu$ is nondynamical,
and the variations with respect to it need not vanish,
\beq
\fr {\partial {\cal L}} {\partial \bar k^\mu}  \ne 0 .
\label{barknondyn}
\eeq
Note that the background $\bar k^\mu$ is different from a current $J^\mu$ 
that is carried by dynamical charged particles or matter fields.
With physical charge-carrying fields, the variations in \rf{EMNoether}
would be with respect to those fields, 
and the corresponding Euler-Lagrange expressions would vanish on shell.

Here, however, the vanishing of the last two terms in \rf{EMNoether} becomes
a constraint that the metric or vector field must satisfy
in addition to their equations of motion.
With $\fr {\partial {\cal L}} {\partial \bar k^\mu} = A_\mu$,
the on-shell consistency conditions become
\beq
(D_\mu A_\nu)\bar k^\mu + A_\mu (D_\nu \bar k^\mu) = 0 .
\label{EMcond1}
\eeq
Writing out the covariant derivatives shows that the metric drops out,
and the conditions reduce to
\beq
(\partial_\mu A_\nu) \bar k^\mu + A_\mu( \partial_\nu \bar k^\mu) = 0 .
\label{EMcond2}
\eeq
The vector $A_\mu$ by itself must satisfy these conditions,
and the result is therefore highly constrained and does not permit
generic dynamical interactions.
For example,
if $\bar k^\mu = \rh (t) \de^\mu_0$,
the vector potential must obey
$\rh \partial_0 A_\nu + A_0 \partial_\nu \rh = 0$,
which no longer allows any independent dynamical 
degrees of freedom in $A_\mu$.

Note that in addition to the Noether identity associated with GCI,
consistency with the Maxwell equations, $D_\mu F^{\mu\nu} = -\bar k^\nu$,
also requires that $D_\mu \bar k^\mu = 0$ must hold.
This puts additional constraints on the theory,
since the current conservation is not the result of
dynamical matter equations of motion.
Instead, the vanishing covariant divergence requires that
$\partial_\mu (\sqrt{-g} \bar k^\mu ) = 0$ must hold,
which either restricts the spacetime geometry
or requires that $\bar k^\mu$ must vanish.
It is for this reason that the SME does not include
terms of the form $\bar k^\mu A_\mu$ in the gravity-photon sector
even with spontaneous symmetry breaking
\cite{akgrav04}.

As this example illustrates,
adding dynamical matter fields that can interact with the
nondynamical background does not necessarily ease
the consistency conditions imposed by Noether's identity.
In fact, it can have the opposite effect of requiring 
additional constraints that need to be satisfied.

\subsection{Massive Gravity}

Interest in theories of massive gravity has increased substantially in recent years,
since it was demonstrated that the models of de Rham, Gabadadze, and Tolley (dRGT)
\cite{MGreviews,dRGT,HR}
do not contain a Boulware-Deser ghost
\cite{BD72}.
The dRGT models can therefore be used to describe physical massive gravitons
in the context of effective field theory.

The mass terms in models of massive gravity are constructed using potentials,
${\cal L}(g_{\mu\nu},\bar k_{\mu\nu})$,
which couple the metric with a symmetric two-tensor background,
obeying $\bar k_{\mu\nu} = \bar k_{\nu\mu}$.
The nondynamical background is needed to generate mass terms for $g_{\mu\nu}$,
since quadratic and higher-order products cannot be formed for the metric by itself.
In the original versions, a Minkowski background,
$\bar k_{\mu\nu} = \et_{\mu\nu}$ was used,
but it has also been shown that ghost-free models with
a more general background $\bar k_{\mu\nu}$ exist as well.

There are several formulations of massive gravity models,
including both metric and vierbein formalisms.
It is common to use a St\"uckelberg approach in
theories of massive gravity.
However, these are equivalent to the theories with nondynamical backgrounds
and will not be considered separately here.  

In a metric formalism,
the dRGT models are defined in terms of square roots of
the inverse metric contracted with the background,
which can be written as
\beq
\sqrt{g^{\mu\al} \bar k_{\al\nu}} = \left( \sqrt{g^{-1} \bar k} \right)^\mu_{\pt{\mu}\nu} .
\label{gasqrt1}
\eeq
The generic form of the
Noether identity associated with GCI then has the form
\bea
g_{\nu\al} D_\mu (G^{\mu\nu} - T^{\mu\nu}) 
\quad\quad\quad\quad\quad\quad\quad\quad\quad\quad
\nonumber \\
- \, 2 D_\mu \left( \fr {\partial {\cal L}} { \partial \bar k_{\mu\nu}} \bar k_{\al \nu} \right)
+ \fr {\partial {\cal L}} { \partial \bar k_{\mu\nu}} D_\al \bar k_{\mu\nu} = 0 .
\label{dRGTNoether}
\eea
Consistency in this case requires that the metric must take values that 
set the last two terms to zero on shell.
Only then can covariant energy-momentum conservation hold.

The presence of square roots in the action 
requires that the field variations must be handled with care
\cite{Volk12}.
The question of whether the square-root matrices exist is relevant as well.
However, in many calculations,
it is either assumed that the square-root matrices exist or
an antsatz form is used that permits the square root to be found.
In the latter case, however, if not enough degrees of freedom
are included for the metric,
this can lead to inconsistency when no solutions
to the Noether identity are found.
For example,
when a Minkowski background field is used,
so that $\bar k_{\mu\nu} = \et_{\mu\nu}$,
and the physical metric $g_{\mu\nu}$ is assumed to be spatially flat and
homogeneous and isotropic,
then there is no exact solution in dRGT gravity
\cite{AdRDGPT11}.
Here, this can be understood by the fact that with the
assumptions being made,
there are not enough degrees of freedom in the Noether identity
\rf{dRGTNoether} to make the last two terms vanish.
However, when an alternative form for the background is used
besides $\et_{\mu\nu}$,
which introduces additional components in the constraint equations,
then an exact solution describing a spatially flat homogeneous
and isotropic universe has been obtained
\cite{GLM}.

Using a vierbein as defined in \rf{gee} introduces a natural square root for the metric.
Similarly, the symmetric two-tensor background $\bar k_{\mu\nu}$ can be written
in terms of the background vierbein as
\beq
\bar k_{\mu\nu} = \barvb \mu a \barvb \nu b \bar k_{ab} ,
\label{kmunukab}
\eeq
where $\bar k_{ab} = \bar k_{ba}$ are the components of the background
in the local frame.

The Noether identity stemming from observer LLI can then be used to look for
constraints that the vierbein must satisfy,
which can then shed light on the nature 
of the square-root matrices in dRGT models of massive gravity.

With the Lagrangian potential having the dependent form
${\cal L}(\vb \mu a,\barvb \mu a,\bar k_{ab})$,
the resulting Noether identity is
\bea
(G^{\mu\nu} - T^{\mu\nu}) (\lvb \mu a \lvb \nu b - \lvb \mu b \lvb \nu a)
+ (\fr {\de {\cal L}} {\de \barvb \mu a} \barlvb \mu b - \fr {\de {\cal L}} {\de \barvb \mu b} \barlvb \mu a )
\nonumber
\\
+ 2 \fr {\de {\cal L}} {\de \bar k_{cd}} \left( \et_{ac} \bar k_{bd} - \et_{bc} \bar k_{ad} \right) = 0 .
\quad\quad\quad
\label{vierLLINoethermassive1}
\eea
However,
for Lagrangians where the functional dependence on $\vb \mu a$, $\barvb \mu a$, and $\bar k_{ab}$ 
occurs only through the combinations ${\cal L}(g_{\mu\nu},\bar k_{\mu\nu})$,
the resulting Noether identity can be shown to reduce to a trivial identity 
that does not impose a constraint on the physical vierbein.
This is because with ${\cal L}$ depending on the combination $\bar k_{\mu\nu} = \barvb \mu a \barvb \nu b \bar k_{ab}$,
it follows that
\beq
\fr {\de {\cal L}} {\de \barvb \mu a} \barlvb \mu b = 2  \fr {\de {\cal L}} {\de \bar k_{cd}}\et_{bc} \bar k_{ad} ,
\label{kident}\eeq
and therefore there are cancelations in \rf{vierLLINoethermassive1}
that reduce the identity to simply $T^{\mu\nu} = T^{\nu\mu}$.

In massive gravity, 
it is common to use a redefinition of the background vierbein
that is unique for the case of a symmetric two-tensor.
The implications of Noether's identity stemming from observer LLI
can be investigated in this context as well.

In this approach,
instead of using $\barvb \mu a$ which links the components of the background in
the spacetime and local frames,
a different background vierbein $\bar v_\mu^{\pt{\mu} a}$ is introduced
\cite{hr12,dmz}.
It is defined by
\beq
\bar k_{\mu\nu} = \bar v_\mu^{\pt{\mu} a} \bar v_\nu^{\pt{\mu} b} \et_{ab} .
\label{vvier}
\eeq
In this case, 
the redefined background vierbein $\bar v_\mu^{\pt{\mu} a}$ gives $\bar k_{\mu\nu}$ 
in terms of the same local Minkowski background $\et_{ab}$ that the metric
equals in the local frame.

Notice that substituting $\bar v_\mu^{\pt{\mu} a}$ and $\et_{ab}$,
respectively, for $\barvb \mu a$ and $\bar k_{ab}$ in \rf{vierLLINoethermassive1}
results in a simplified Noether identity.
This is because the last set of terms in \rf{vierLLINoethermassive1} with $\bar k_{ab} = \et_{ab}$
vanishes identically,
and therefore the Noether identity reduces to
\bea
(G^{\mu\nu} - T^{\mu\nu}) (\lvb \mu a \lvb \nu b - \lvb \mu b \lvb \nu a)
\quad\quad\quad\quad
\nonumber
\\
+ (\fr {\de {\cal L}} {\de \bar v_\mu^{\pt{\mu} a}} \bar v_{\mu b} 
- \fr {\de {\cal L}} {\bar v_\mu^{\pt{\mu} b}} \bar v_{\mu a} )
= 0 .
\quad\quad\quad
\label{vierLLINoethermassive2}
\eea

However, even in this form, as long as the Lagrangian still
has the dependence ${\cal L}(g_{\mu\nu},\bar k_{\mu\nu})$,
with $\bar k_{\mu\nu}$ now given by \rf{vvier},
then a trivial Noether identity follows.
In this case it is because 
\beq
\fr {\de {\cal L}} {\de \bar v_\mu^{\pt{\mu} a}} \bar v_{\mu b}
= \fr {\de {\cal L}} {\de \bar v_\mu^{\pt{\mu} b}} \bar v_{\mu a}
\label{var0}
\eeq
holds identically due to the form of the dependence.

To obtain a nontrivial Noether identity due to observer LLI in dRGT massive gravity,
a Lagrangian potential that is not invariant under particle local Lorentz transformations
must be used so that there is coupling between the physical and background vierbeins.
An approach that is used in dRGT theories involves a matrix defined as
\beq
\ga^\mu_{\pt{\mu}\nu} = \ivb \mu a \bar v_\nu^{\pt{\nu}a} ,
\label{gasqrt}
\eeq
which is not invariant under particle local Lorentz transformations.
In a vierbein formulation,
the action can then be written in terms of this matrix,
where the Lagrangian is formed as observer scalar combinations of $\ga^\mu_{\pt{\mu}\nu}$.
In four dimensions there are four independent observer scalars that can be formed
as traces of products of the matrix $\ga^\mu_{\pt{\mu}\nu}$.
These are denoted as
\beq
X_n = {\rm tr}  [ \ga^n ] , 
\label{Xn}
\eeq
with $n = 1,2,3,4$.
The functional dependence of the Lagrangian potential can then be given as ${\cal L}(X_n)$.

With these definitions,
the gravitational action can be written as
\beq
S= \int d^4x \sqrt{-g} \left[ \fr 1 2 R + {\cal L} (X_n) \right] .
\label{SdRGTr}
\eeq
Matter terms can be included as well.
However, these do not have interactions with the nondynamical background,
and for this reason they are not included here.

The Noether identity stemming from observer LLI in the
action \rf{SdRGTr} then has the form
\bea
(G^{\mu\nu} - T^{\mu\nu}) (\lvb \mu a \lvb \nu b - \lvb \mu b \lvb \nu a)
\quad\quad\quad\quad\quad\quad
\nonumber
\\
+ \sum_{n=1}^4 \fr {\partial {\cal L}} {\partial X_n} 
\left(\fr {\partial X_n} {\partial \bar v_\mu^{\pt{\mu}a} } \bar v_{\mu b} 
- \fr {\partial {X_n}} {\partial \bar v_\mu^{\pt{\mu}b} } \bar v_{\mu a} \right)
= 0 .
\quad
\label{vierLLINoethermassive}
\eea
In this case, the identity is not trivial due to the coupling
between $\vb \mu a$ and $\bar v_\mu^{\pt{\mu}a}$,
and constraints are imposed on the physical vierbein.

When the Einstein equations hold on shell,
consistency requires that the sum in \rf{vierLLINoethermassive} must vanish.
Depending on which scalars $X_n$ are included in the action,
there can be multiple solutions or branches that satisfy the consistency conditions
\cite{dmz}.
However, a sufficient condition that results in the sum vanishing is
that the physical vierbein must obey a symmetry condition,
\beq
\ivb \mu a  \bar v_{\mu b} - \ivb \mu b \bar v_{\mu a} = 0 .
\label{symv}
\eeq
Notice that this symmetry condition \rf{symv} has the same form as
the Deser-van Nieuwenhuizen gauge-fixing condition in \rf{syme},
except that here the backgroud vierbein $\bar v^\mu_{\pt{\mu} a}$
appears in place of $\barvb \mu a$.

It is also the case that when \rf{symv} holds,
then the square of $\ga^\mu_{\pt{\mu}\nu}$ as defined in \rf{gasqrt} obeys
\beq
\ga^\mu_{\pt{\mu}\al} \ga^\al_{\pt{\mu}\nu} = g^{\mu\al} \bar k_{\al\nu} .
\label{gasquared}
\eeq
This therefore verifies the existence of the square-root matrix for the metric,
\beq
\sqrt{g^{\mu\al} \bar k_{\al\nu}} = \ga^\mu_{\pt{\mu}\nu} ,
\label{gasqrt2}
\eeq
when the symmetry condition \rf{symv} holds.

While the metric formulation of dRGT does not have a Noether identity
stemming from observer LLI,
the vierbein formulation in terms of $\ga^\mu_{\pt{\mu}\nu}$ does,
and it imposes the conditions \rf{symv} on the physical vierbein.
Evidently, when these conditions are satisfied,
as found originally in \cite{hr12,dmz},
this also establishes the existence of a square-root matrix for the metric.
Thus, a linkage is formed between the metric and vierbein formalisms
when the symmetry condition stemming from Noether's identity is satisfied.

The effect of the observer LLI Noether identity in
massive gravity defined using the symmetry-breaking matrices $\ga^\mu_{\pt{\mu}\nu}$
is that it imposes consistency conditions
on the physical vierbein that makes the nondynamical vierbein compatible
with the choice of local frame.
Once compatibility is assured,
then the physical and background vierbeins can be combined to
give both $g_{\mu\nu}$ and $\bar k_{\mu\nu}$.
In the process, the antisymmetric components of the physical and background vierbeins drop out.
However, those components are necessary in establishing that the square-root matrix
in the metric formalism actually exists.
In this way, the Noether identity due to observer LLI ends up 
playing a key role in the metric formulation.

\section{Summary and Conclusions}

Gravity theories with fixed background fields break diffeomorphism
invariance and LLI either explicitly or spontaneously.
In the case of spontaneous breaking,
the background fields arise as dynamical vacuum solutions,
$\bar k_{\la\mu\nu\cdots}$, which satisfy vacuum Euler-Lagrange equations.
The interpretation of the Noether identities in theories with spontaneous spacetime
symmetry breaking is therefore the same as in GR.

In contrast, when a gravitational theory contains a nondynamical background field, 
diffeomorphism invariance and LLI are explicitly broken, 
and a different form and interpretation of Noether identities emerges. 
First, a distinction must be made between the broken particle symmetries 
and the mathematical observer transformations that leave the action invariant. 
In particular, the observer invariances are required if a theory 
is to maintain observer independence. 
Therefore, gravity theories with a nondynamical background must still 
have local GCI and observer LLI as mathematical invariances in the action. 
These mathematical observer invariances can then be used to obtain Noether identities 
that hold even when there is explicit spacetime symmetry breaking. 
However, an important feature in this context 
is that the background fields do not need to satisfy 
their Euler-Lagrange equations.
This makes the Noether identities associated with observer independence
very different from the usual case in GR.

The loss of diffeomorphism invariance and LLI also means
that there are additional metric and vierbein degrees of freedom 
that appear in Einstein's equations in comparison to GR.
These are the degrees of freedom that would normally be gauged away in GR
or in theories with spontaneous spacetime symmetry breaking.
It is these extra metric and vierbein modes that have to satisfy the Noether identities
when there is explicit breaking.
They also must absorb the backreactions that the immoveable background field is unable to have,
which then allows covariant energy-momentum conservation to hold.

When the nondynamical background tensor is given specified values 
$\bar k_{\la\mu\nu\cdots}$ with respect to a particular coordinate frame
as well as specified values $\bar k_{abc\cdots}$ with respect to a
particular local Lorentz basis,
the observer invariances become fixed.
The consistency of the theory then requires that there must be compatibility
between the form of the background,
the choice of coordinates and local Lorentz frame,
and the solutions for the metric and vierbein.
It is the Noether identities associated with the local observer invariances
that give these compatibility conditions.

In certain examples,
the conditions imposed by Noether's identities match known gauge-fixing conditions in GR.
However, this does not mean that the solutions are equivalent to solutions in GR,
since the changes in the constraint structure can also result in additional degrees of
freedom that are physical.
Indeed, much of the interest in including background fields in gravity 
is aimed at finding modified theories with additional physical modes
that can provide alternative approaches to solving issues related to
quantum gravity, dark energy, or dark matter.

Lastly, in some cases,
the Noether identities can rule out a theory completely or place restrictions 
on the geometry or matter dynamics
\cite{fins}.
In this way, they can provide a useful tool that can be 
used to investigate gravity theories with 
nondynamical backgrounds and explicit spacetime symmetry breaking.

%\section*{Acknowledgments}

\end{document}